# SUSTAINED COMOVING ACCELERATION OF OVERDENSE PLASMAS BY COLLIDING LASER PULSES

EDISON LIANG

*Rice University, Hsouton, TX 77005-1892 USA*

We review recent PIC simulation results which show that double-sided irradiation of a thin over-dense plasma slab by ultra-intense laser pulses from both sides can lead to sustained comoving acceleration of surface electrons to energies much higher than the conventional ponderomotive limit. The acceleration stops only when the electrons drift transversely out of the laser beam. We show latest 2.5D results of parameter studies based on finite laser spot size and discuss future laser experiments that can be used to test these results.

## 1. Introduction

Recent advances in ultra-intense short-pulse lasers (ULs) [1,2] open up new frontiers on particle acceleration via ultra-strong electromagnetic (EM) fields [3]. Most conventional laser acceleration schemes (e.g. LWFA, PWFA, PBWA [4], FWA [5]) involve the propagation of lasers in an underdense plasma ($\omega_{pe}=(4\pi n e^2/m_e)^{1/2}<\omega_o=2\pi c/\lambda$, $\lambda$=laser wavelength, n=electron density). In such schemes the acceleration gradient (energy gain/distance) [4] and energetic particle beam intensity are limited by the laser frequency due to the underdense requirement. Here we review PIC simulation results of a radically different concept: comoving acceleration of overdense ($\omega_{pe}>\omega_o$) plasmas using colliding UL pulses. In this case the acceleration gradient and particle beam intensity are not limited by the underdensity condition. This colliding pulses accelerator (CLPA) mechanism may have important applications complementary to those of underdense laser acceleration schemes.

Consider an intense EM pulse with $\Omega_e(=a_o\omega_o=eB_o/m_e c$, $a_o$=normalized vector potential)$>\omega_{pe}$ initially imbedded inside an overdense plasma ($\omega_{pe}>>\omega_o$). When it tries to escape, it induces a diamagnetic skin current J that inhibits the EM field from leaving. The resultant **J x B** (ponderomotive) force then accelerates the surface plasma to follow the EM pulse. As the EM pulse "pulls" the surface plasma, it is slowed by plasma loading (group velocity < c), allowing the fastest particles to comove with the EM field. But since slower particles eventually fall behind, the plasma loading decreases and the EM pulse accelerates with time. A dwindling number of fast particles also get accelerated indefinitely by the comoving EM force, reaching maximum Lorentz factors greater than the usual ponderomotive limit [6] $\gamma_{max}> a_o^2/2 >>(\Omega_e/\omega_{pe})^2$. This novel phenomenon is called the diamagnetic relativistic pulse accelerator (DRPA) [7]. DRPA is strictly a nonlinear, collective, relativistic phenomenon, with no analog in the weak field ($\Omega_e/\omega_{pe}<1$), low density ($\omega_o>\omega_{pe}$) or test particle limit. Here we discuss a laser acceleration scheme based on the DRPA concept.

## 2. Colliding Pulses Acceleration Mechanism

Since the discovery of DRPA from PIC simulations, a key question has been how to reproduce it in the laboratory, as vacuum EM waves cannot penetrate an overdense plasma beyond the relativistic skin depth [8]. Fig.1 shows the PIC simulation of a single UL irradiating an overdense e+e- plasma. All upstream plasma is snowplowed by the UL, and the terminal maximum Lorentz factor $\gamma_{max} \sim (\Omega_e/\omega_{pe})^2$. The relativistic mass increase [8] is countered by density increase due to compression, and the plasma stays overdense at all times, preventing the UL from penetrating. Hence the DRPA initial condition cannot be achieved using a single UL pulse. Here we report PIC simulations with the 2.5D (2D-space, 3-momenta) ZOHAR code [9], which demonstrate that DRPA-like sustained comoving acceleration can be achieved by irradiating a thin slab of overdense e+e- plasma with UL pulses from opposite sides. The opposing UL pulses accomplish this by first compressing the overdense plasma to a total thickness < 2 relativistic skin depths [8]. At that point the UL pulses "tunnel" through the plasma, despite its overdensity even allowing for relativistic effect ($\omega_{pe} > <\gamma>\omega_o$, $<\gamma>$=mean Lorentz factor of the compressed plasma). The physics of the subsequent evolution after transmission is similar to that of the DRPA [7].

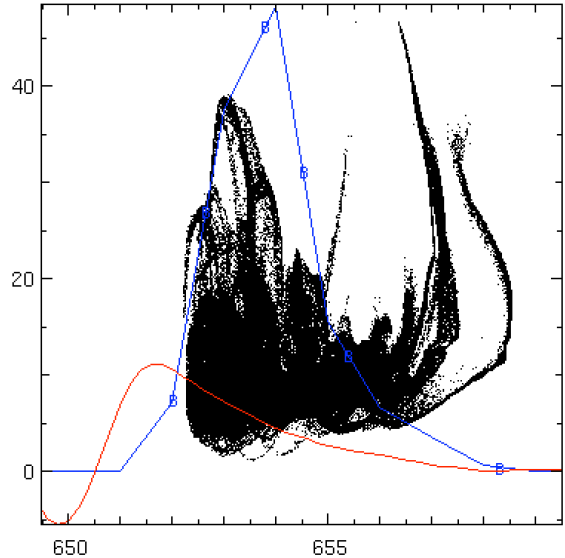

Fig.1. PIC simulation shows that a single UL pulse ($I(\lambda/\mu m)^2=10^{21}W/cm^2$, $c\tau=\lambda/2$) snowplows an overdense ($n_o=15n_{cr}$, thickness = $\lambda/2$, kT=2.6keV) e+e- plasma but cannot penetrate it. We plot $B_y$; $n/n_{cr}$ (B) and $p_x/mc$ (black dots) vs. x at $t\omega_o/2\pi = 20$. The slab thickness remains >> relativistic skin depth at all times. The maximum Lorentz factor $\gamma_{max} \sim (\Omega_e/\omega_{pe})^2 \sim 40$ at late times [16].

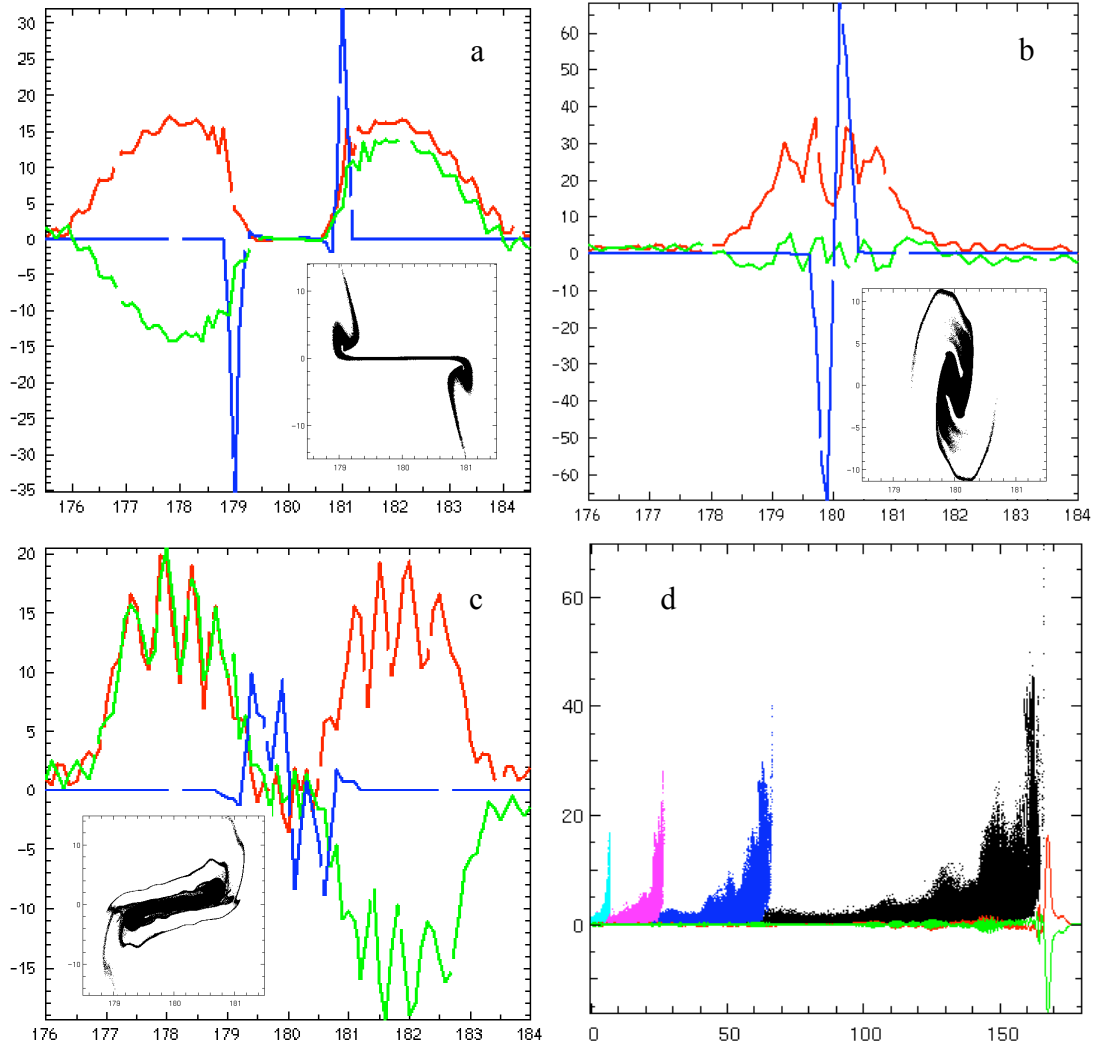

Fig.2.. Evolution of two linearly polarized plane EM pulses ($I(\lambda/\mu m)^2=10^{21}W/cm^2$, $c\tau=\lambda/2$) irradiating an overdense e+e- plasma ($n_o=15n_{cr}$, thickness = $\lambda/2$, kT=2.6keV) from opposite sides. We plot magnetic field $B_y$(medium), electric field $E_z$(light), current density $J_z$(dark) and $p_x/mc$ vs. x (inset) at $t\omega_o/2\pi$ = (a)1.25, (b)1.5, (c)1.75; (d) Snapshots of $p_x/m_ec$ vs. x (dots) for the right-moving pulse at $t\omega_o/2\pi$=2.5(black), 5(red), 10(blue), 22.5(green) showing power law growth of $\gamma_{max}\sim t^{0.45}$. We also show the profiles of $B_y$(medium), $E_z$(light) at $t\omega_o/2\pi$=22.5. [16]

Fig.2 shows the evolution of two linearly polarized plane half-cycle EM pulses with parallel **B**, irradiating a thin e+e- slab from opposite sides (thickness=$\lambda/2$, initial density $n_o=15n_{cr}$(critical density)). Cases with nonparallel **B** are more complex and are still under investigation. Each incident pulse compresses and accelerates the plasma inward (Fig.2a), reaching a terminal Lorentz factor $\gamma_{max}\sim(\Omega_e/\omega_{pe})^2\sim 40$ as in Fig.1. Only ~10% of the incident EM amplitudes is reflected because the laser reflection front is propagating inward relativistically [10]. As the relativistic skin depths from both sides start to merge (Fig.2b), the two UL pulses interpenetrate and tunnel through the plasma, despite $\omega_{pe} > <\gamma>^{1/2}\omega_o$. Such transmission of EM waves through an overdense plasma could not be achieved using a single UL pulse, because there the plasma thickness remains $\gg$ 2 relativistic skin depths (Fig.1). During transmission, the **B** fields of the opposing pulses add while **E** fields cancel (Fig.2b), setting up a state similar to the DRPA initial state, and the subsequent evolution resembles the DRPA [7]. As the transmitted UL pulses reemerge from the plasma, they induce new drift currents **J** at the *trailing* edge of the pulses (Fig.2c), with opposite signs to the initial currents (Fig.2b), so that the new **J x B** forces pull the surface plasmas outward. We emphasize that the plasma loading which slows the transmitted UL pulses plays a crucial role in sustaining this comoving acceleration. As we see in the parameter study below, for a given $\Omega_e/\omega_{pe}$ the higher the plasma density, the more sustained the comoving acceleration, and a larger fraction of the plasma slab is accelerated. This unique feature distinguishes this overdense acceleration scheme from other underdense schemes [4,5]. As slower particles gradually fall behind the UL pulses, the plasma loading of the UL pulses decreases with time. This leads to continuous acceleration of both the UL pulses and the dwindling population of trapped fast particles. The phase space evolution (Fig.2d) of this colliding pulses accelerator (CLPA) resembles that of the DRPA [7].

## 3. Acceleration by Gaussian Pulse Trains

The above example using half-cycle UL pulses is only for illustration. Fig.3 shows the results of irradiating an overdense e+e- slab using more realistic Gaussian pulse trains ($\lambda=1\mu m$, pulse length $\tau=85fs$, $I_{peak}=10^{21} Wcm^{-2}$). We see that $\gamma_{max}$ increases rapidly to 2200 by 1.28ps and 3500 by 2.56ps, far exceeding the ponderomotive limit $a_o^2/2$ (~360). The maximum Lorentz factor increases with time according to $\gamma_{max}(t) \sim e\int E(t)dt/mc$. $E(t)$ is the UL electric field comoving with the highest energy particles. $E(t)$ decreases with time due to EM energy transfer to the particles, plus slow dephasing of particles from the UL pulse peak. This leads to $\gamma_{max}$ growth slower than linear and $\gamma_{max} \sim t^{0.8}$ (Fig.3b). In practice, $\gamma_{max}$ will be limited by the diameter D of the laser focal spot, since particles drift transversely out of the laser field after $t\sim D/c$. The maximum energy of any comoving acceleration is thus $< eE_oD = 6GeV(I/10^{21}Wcm^{-2})^{1/2}(D/100\mu m)$. The asymptotic momentum distribution forms a power-law with slope $\sim -1$ (Fig.3d) below $\gamma_{max}$, distinct from the exponential distribution of ponderomotive heating [11,12]. We speculate that a quasi-power-law momentum distribution is formed below $\gamma_{max}$ since there is no other preferred energy scale below $\gamma_{max}$, and the particles have random phases with respect to the EM field profile.

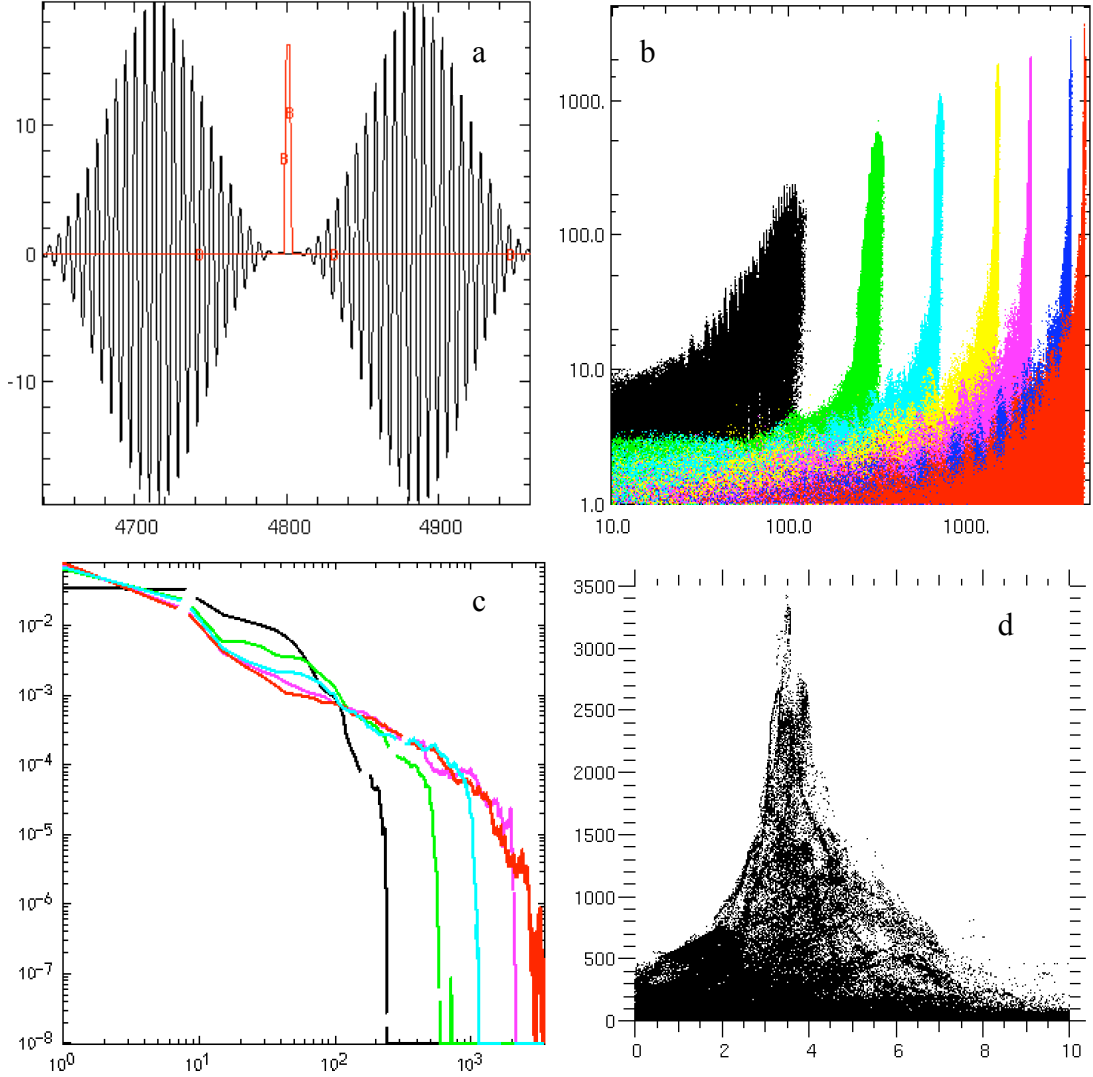

Fig.3. Results of two Gaussian pulse trains ($\lambda=1\mu m$, $I=10^{21}W/cm^2$, $c\tau=85fs$) irradiating a e+e- plasma ($n_o=9n_{cr}$, thickness = $2\lambda/\pi$, $kT=2.6keV$). (a) early $B_y$ and $n_o/n_{cr}$ (B) profiles at $t\omega_o=0$; (b) time-lapse evolution of $\log(p_x/m_ec)$ vs. $\log x$ for the right-moving pulse at $t\omega_o=$ (left to right) 180, 400, 800, 1600, 2400, 4000, 4800 showing power-law growth of $\gamma_{max}\sim t^{0.8}$; (c) evolution of electron energy distribution $f(\gamma)$ vs. $\gamma$ showing the build-up of power-law below $\gamma_{max}$ with slope $\sim -1$: $t\omega_o=$ (left to right) 180, 400, 800, 2400, 4800. (Slope $=-1$ means equal number of particles per decade of energy), (d) plot of $\gamma$ vs. $\theta$ ($=|p_z|/|p_x|$) in degrees at $t\omega_o=4800$, showing strong energy-angle selectivity and narrow beaming of the most energetic particles. [16]

## 4. Parameter Studies

We have performed extensive parameter studies of the CLPA. Since $\gamma_{max}$ is not the only figure of merit in comparing acceleration efficiency here we compare the overall particle energy distributions at equal times for different runs. Fig4a shows the effects of varying vector potential $a_o$ while fixing other parameters. Both the power-law slope and $\gamma_{max}$ increase with $a_o$. Fig4b shows the effect of increasing the pulse length $\tau$ while fixing other parameters. We see that at first $\gamma_{max}$ increases and the power-law slope stays ~constant, but for very long pulses $\gamma_{max}$ becomes fixed while the slope hardens. Fig4c shows the effect of varying the target density n while fixing other parameters. For comparison we include the result of an underdense example ($n=10^{-3}n_{cr}$ bottom curve). While all three cases produce similar $\gamma_{max}$, the underdense case shows a smaller fraction of particles being accelerated, because the plasma loading is too low to effectively slow down the UL pulses. The physics of the underdense CLPA may be related to the free wave accelerator (FWA [5]), where we substitute the symmertry-breaking electrostatic/magnetostatic field [5] with an opposing laser. But as Fig.4c shows, the overdense CLPA is more effective in terms of energy coupling and the fraction of plasma accelerated.

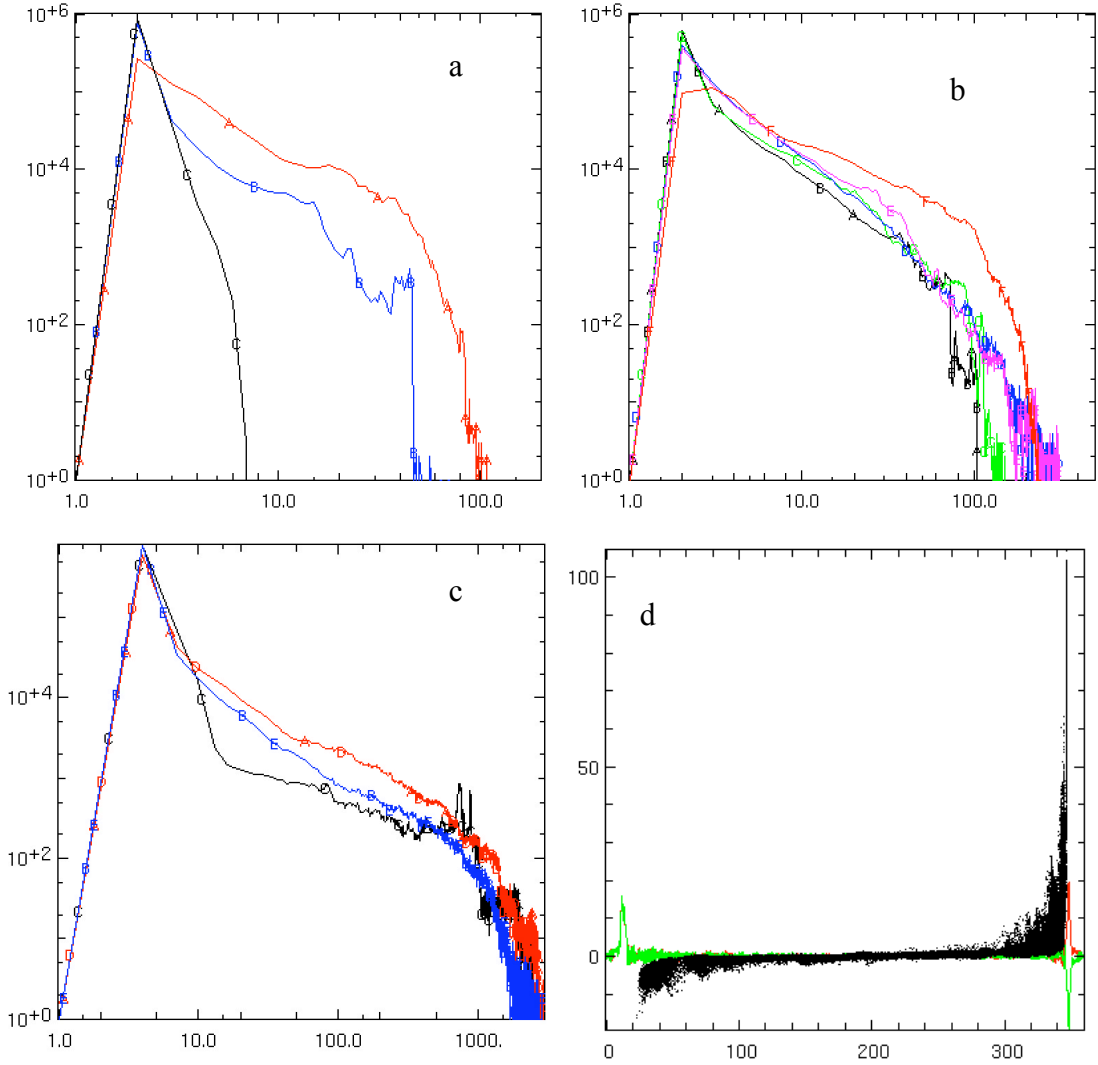

Fig.4. Comparison of electron energy distribution $f(\gamma)$ vs. $\gamma$ at equal times when we vary a single input parameter:(a) $a_o$=1.9,19,190 at $t\omega_o/2\pi$=22.5; (b) $c\tau$= $\lambda/2$, $\lambda$, $4\lambda$, $7\lambda$, $26\lambda$ at $t\omega_o/2\pi$=22.5; (c) $n_o/n_{cr}$= 9, 25, 0.001 at $t\omega_o$=4800; (d) phase plots (dots), magnetic field $B_y$, (medium) electric field $E_z$ (light) at $t\omega_o/2\pi=22.5$ when two unequal UL pulses (see text for laser intensities) irradiate the same plasma as in Fig.2. [16]

We have also studied the effects of unequal intensities from the opposing laser pulses. Fig.4d illustrates the case in which a thin plasma slab is irradiated by a UL of $10^{21}$ Wcm$^{-2}$ from the right and $8\times10^{20}$ Wcm$^{-2}$ from the left. We see that most of the particles are trapped and accelerated by the right-moving pulse while the left-moving pulse decouples from the plasma early, leading to little trapping or acceleration.

## 5. Proposed Laser Experiment

An experimental demonstration of the CLPA will require a dense and intense e+e- source. Cowan et al [13] demonstrated that such an e+e- source can be achieved by using a PW laser striking a gold foil. Theoretical works [14] suggest that e+e- densities $>10^{22}$cm$^{-3}$ may be achievable with sufficient laser fluence. Such a high density e+e- jet can be slit-collimated to produce a ~ micron thick e+e- slab, followed by 2-sided irradiation with opposite UL pulses. As an example, consider UL pulses with $\tau$=80fs and intensity=$10^{19}$Wcm$^{-2}$. We need focal spot diameter D>600 µm for the pairs to remain inside the beam for >1ps. This translates into ~1KJ energy per UL pulse. Such high-energy UL's are currently under construction at many sites [2]. Fig.5 shows the artist conception of such an experiment setup.

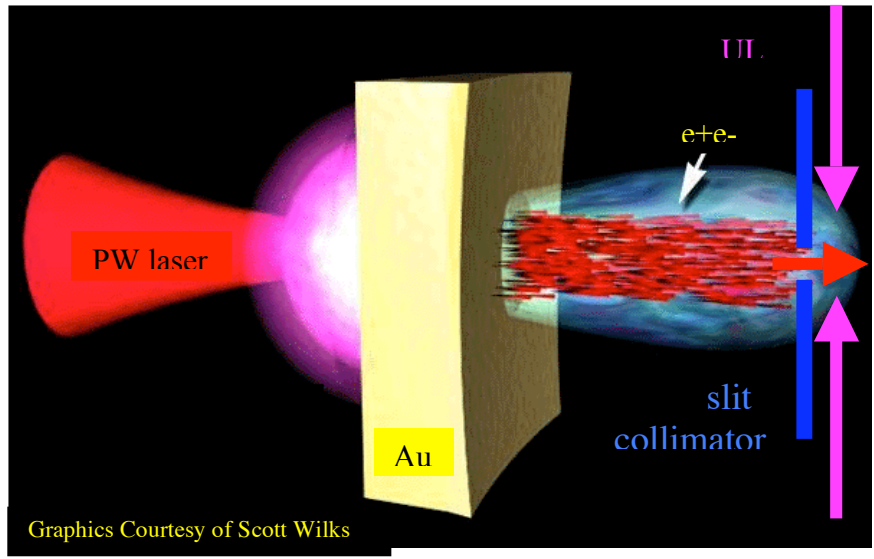

Fig.5. Conceptual experiment setup for the demonstration of the CLPA mechanism using three PW lasers.

## 6. Electron-ion Plasmas

We have also begun investigating the CLPA concept for e-ion plasmas. Preliminary results suggest that, for very thin e-ion plasma slabs which can be compressed to <

two relativistic skin depths, the CLPA concept remains viable. Most energy is eventually transferred to ions via charge separation, similar to the e-ion DRPA [7,15]. Fig.6 compares CLPA runs for e-ion plasmas at different ion densities. We see that the higher the ion density, the stronger the charge-separation electric fields dragging the electrons. So there must be a trade-off between plasma loading which slows the UL pulses, and the ion drag which slows the electron component. Details remain to be investigated. Though the previous results reported are based on 2.5D simulations, new 3D results confirm the stability and robustness of the CLPA concept.

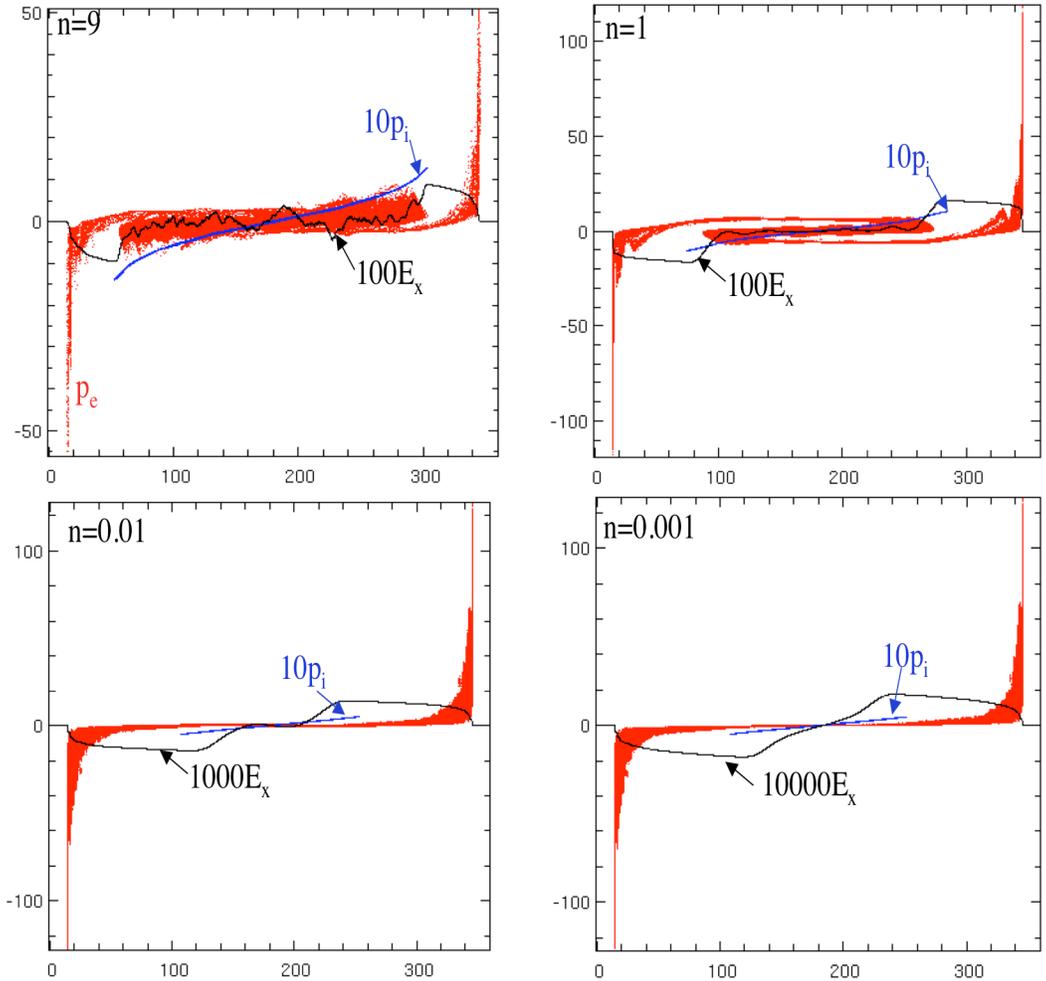

Fig.6. Electon and ion phase plots and charge separation electric field profiles, when a thin slab (thickness=$\lambda/2$) of e-ion plasma is irradiated from both sides by the same UL pulses as in Fig.2. At high densities most of the energy is transferred to the ions via charge separation and the ion drag on electrons allows only a small fraction of electrons to comove with the UL pulses. At low densities ion acceleration is negligible and most of the electrons are freely accelerated as in the e+e- case. Critical density is equal to 0.6 in these units. [17]

## 7. Effects of Finite Laser Spot Size

The above results were obtained with infinite linearized plane wave laser pulses, with periodic boundary in y (here B is along z). Using the ZOHAR code [9] we have recently performed 2.5-D simulations with finite laser spot size in the y direction. Fig. 7 gives one sample of such runs with laser spot diameter D=8 microns. We see that despite the finite spot size, the compression (upper left panel) and re-explosion (lower left panel) of the opposing laser pulses are stable at least in 2.5D. Most importantly, the particles along the laser axis are trapped and efficiently accelerated, and the late-time phase diagram (lower middle panel) resembles the infinite plane wave results of Fig.2. We do however observe a small amount of charge separation between the electrons and positrons as evidenced by the slight asymmetry in the space distribution of the particles (upper middle and right panels). However this small charge separation in y-direction, which is expected intuitively due to the finite extent of the laser $E_y$ field, does not seem to affect the longitudinal acceleration in the x direction in any major way. We have also performed electron-ion simulations with finite laser spots and the results are similar to those in Fig.6.

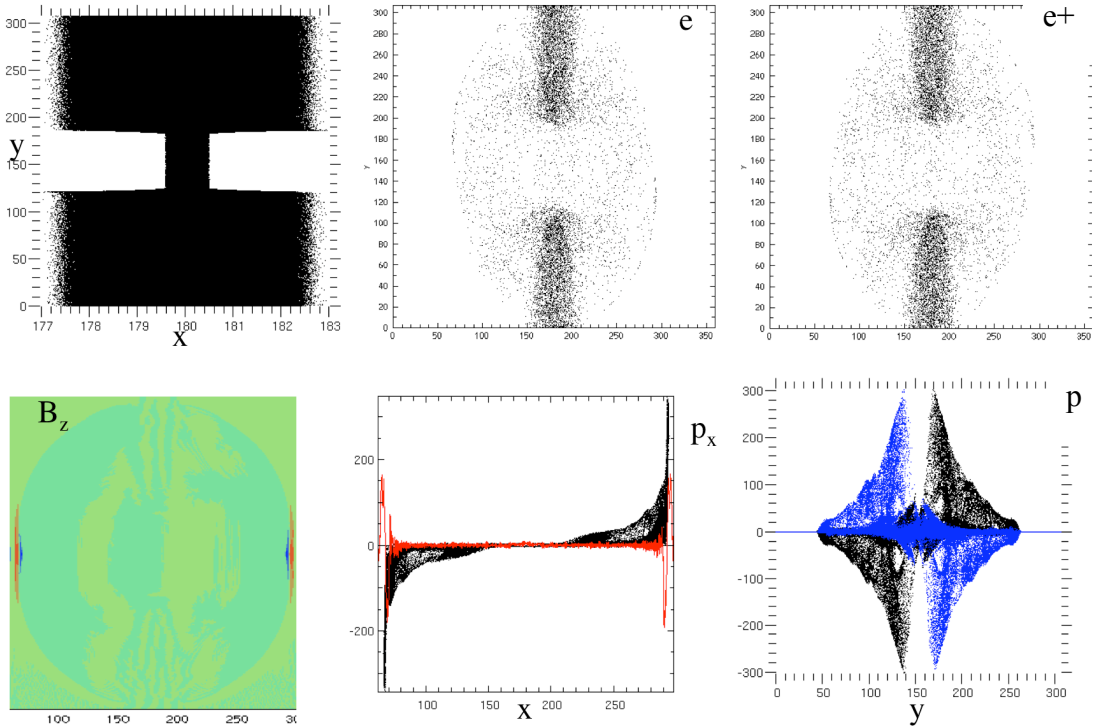

Fig. 7. Snapshots of colliding laser pulses interacting with a central e+e- plasma with finite spot diameter D= 8 microns. Other input parameters are the same as in Fig.2. (upper left) maximum compression of central plasma just before laser pulses tunneling. (upper middle) electron distribution at $t\omega_o$= 140. (upper right) positron distribution at $t\omega_o$=140. (lower left) contour of $B_z$ at $t\omega_o$=140. (lower middle) phase plot of electrons (black dots) and the B field profile (red) along the x-axis at $t\omega_o$=140. (lower right) phase plot along y axis for electrons (blue) and positrons (black) at $t\omega_o$=140 [17].

**Acknowledgement**

EL was partially supported by NASA NAG5-9223, LLNL B537641 and NSF AST-0406882. He acknowledges LANL and LLNL for their support during his sabbatical year when some of these works were performed. He thanks Scott Wilks for helps with running ZOHAR and Fig.5, and Bruce Langdon for providing the ZOHAR code.